\documentclass[a4paper]{aa}
\usepackage{txfonts,graphicx,float}

\def \src {H\thinspace1743$-$322}
\def \degmark{^\circ}

\def \hcm {\hbox {\ifmmode $ atom cm$^{-2}\else atom cm$^{-2}$\fi}}
\def \arcmin {\hbox{$^\prime$}}

\def\approxgt{\mathrel{\hbox{\rlap{\lower.55ex \hbox {$\sim$}}
        \kern-.3em \raise.4ex \hbox{$>$}}}}
\def\approxlt{\mathrel{\hbox{\rlap{\lower.55ex \hbox {$\sim$}}
        \kern-.3em \raise.4ex \hbox{$<$}}}}

\begin{document}

\title{INTEGRAL observations of the black hole candidate 
\src\ in outburst\thanks{Based on observations with INTEGRAL, an ESA 
project with 
instruments and science data centre funded by ESA member states 
(especially the PI countries: Denmark, France, Germany, Italy, Switzerland, 
Spain), Czech Republic and Poland and with the participation of Russia and 
the USA.}}

\author{A. N. Parmar
        \and E. Kuulkers
        \and T. Oosterbroek 
        \and P. Barr
        \and R. Much 
        \and A. Orr
        \and O. R. Williams
        \and C. Winkler
}

\institute{
       Research and Scientific
       Support Department of ESA, ESTEC,
       Postbus 299, NL-2200 AG Noordwijk, The Netherlands
}
\date{Received 11 July 2003 / Accepted 28 July 2003}

\authorrunning{A.N. Parmar et al.}

\offprints{A. N. Parmar, \email{aparmar@esa.int}}

\titlerunning{INTEGRAL observations of \src}

\abstract{
INTEGRAL made 3 observations in 2003 April of the black hole candidate
\src\ during the rising part, and close to the maximum,
of an outburst. \src\ was previously observed in outburst in 1977--1978. 
The source is located in a crowded region of the
sky ($l = 357\degmark, \, b = -2\degmark)$ and at least
18 sources are clearly detected in the field of view of the ISGRI
gamma-ray imager during a 277~ks exposure. These are well known 
persistent X-ray binary sources and 3 transient sources in outburst.
The combined 5--200~keV JEM-X and SPI spectrum of \src\
is well fit with an absorbed 
$((2.5 \, ^{+4.3} _{-2.5}) \times 10^{22}$~atom~cm$^{-2}$) 
soft (photon index $2.70 \pm 0.09$) power-law model, consistent
with \src\ being in a high/soft state.
   
}

\maketitle

\section{Introduction}
\label{sect:intro}

The transient X-ray source \src\ (also referred to as 1H\,1741$-$322)
was discovered during an outburst in
1977--1978 using the HEAO-1 and Ariel~V satellites
(Doxsey et al.~\cite{d:77}; Kaluzienski \& Holt~\cite{kh:77}).
The X-ray spectrum was very soft 
with, at times, a power-law tail extending to $\sim$100~keV
(Cooke et al.~\cite{c:84}). Due to these spectral
characteristics \src\ was proposed as a black hole candidate 
by White \& Marshall~(\cite{wm:84}). 
\src\  was not observed
subsequently, although there was a detection on 1984 August 31, by the
non-imaging EXOSAT Medium Energy instrument during a slew  
maneuver, of a faint ($\sim$5~mCrab) source at a position consistent with 
that of 
\src\ (Reynolds et al.~\cite{r:99}).

On 2003 March 21, when INTEGRAL was performing scans of a region
of sky close to the galactic centre, a bright
source was detected in the field of view of the gamma-ray imager
and designated IGR\,J17464$-$3213 
(Revnivtsev et al.~\cite{re:03}). Subsequent observations showed
that the source flux had increased by a factor 3 by 2003 March 26.
During an RXTE Proportional Counter Array (PCA) scan of the 
galactic bulge region on 2003 March 25, 
Markwardt \& Swank~(\cite{ms:03a}) reported the detection of a
new source which they designated XTE\,J1746$-$322. They 
noted that the source was positionally consistent with IGR\,J17464$-$3213.
PCA observations on 2003 March 28 revealed strong quasi-periodic
oscillations with a period of $\sim$20~s, but no coherent
pulsations were detected. The PCA spectrum is
consistent with a power-law with a photon index, $\alpha$,
of $1.49 \pm 0.01$ and an absorbing column, $N_H$, of 
$2.4 \times 10^{22}$~atom~cm$^{-2}$.
These timing and spectral properties are consistent with
\src\ being a black hole candidate.
The most commonly used position for \src\ is from the HEAO~A-1
catalog (Wood et al.~\cite{w:84}). However, as  
Markwardt \& Swank~(\cite{ms:03b}) point out, the
HEAO~1 Modulation Collimator (MC) instrument determined two
equally probable positions for \src\ (Gursky et al.~\cite{g:78}). 
The HEAO~A-1 catalog reports the western position. The eastern
MC position is consistent with
the INTEGRAL position.
Thus, it is  
highly probable that the source observed by INTEGRAL and RXTE 
in 2003 is indeed \src\ undergoing another outburst.

During the 2003 outburst a compact variable radio source was found 
with the VLA at a 
position consistent with \src\ (Rupen et al.~\cite{ru:03}).
Infrared imaging revealed a likely candidate with a K band magnitude of
$\sim$13--14 (Baba et al.~\cite{b:03}) which may also be visible in I and
R-band images obtained with the Magellan-Clay telescope 
(Steeghs et al.~\cite{s:03}). The best location for 
\src\ is probably the revised VLA position 
 of R.A. = $17{\rm ^h}\,46{\rm ^m}\,15\fs608$,
Decl. = $-32\degmark \, 14\arcmin\ 0\farcs6$ (J2000) 
with an uncertainty of $\pm$$0\farcs5$ (Steeghs et al.~\cite{s:03}).

\section{Observations}
\label{sect:obs}

The INTEGRAL payload (Winkler et al. \cite{w:03})
consists of two gamma-ray instruments, one of which is optimized for 
15~keV to 10~MeV high-resolution
imaging (IBIS; Ubertini et al.~\cite{u:03}) and the other for 
20~keV to 8~MeV high-resolution spectroscopy 
(SPI; Verdrenne et al.~\cite{v:03}). IBIS provides an angular resolution
of $12\arcmin$ full-width at half-maximum (FWHM) and an energy
resolution, $E/\Delta E$, of $\sim$12~FWHM at 100~keV. SPI provides
an angular resolution of $2\fdg 7$~FWHM and an $E/\Delta E$
of $\sim$500~FWHM at 1.3~MeV.  
The extremely
broad energy range of IBIS is covered by two separate detector
arrays, ISGRI (15--500~keV) and PICsIT (0.2--10~MeV).
The payload is completed by X-ray (JEM-X; 3--35~keV; Lund et al.~\cite{l:03})
and optical monitors (OMC; V-band; Mas-Hesse et al.~\cite{m:03}).
The instruments are co-aligned and are operated simultaneously.

\begin{figure}
  \hbox{\hspace{0.2cm}
  \includegraphics[height=8.0cm,angle=-90]{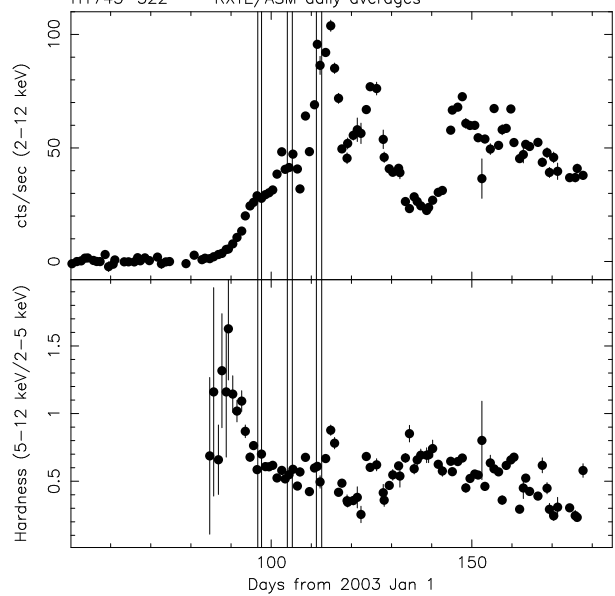}}
  \caption[] {2--12 keV RXTE ASM light curve of part of the 2003
             outburst of \src\ (upper panel). The lower panel shows the
             hardness ratio (counts between 5--12~keV divided by those
             between 2--5~keV). The three pairs of vertical lines indicate
             the start and end times of the INTEGRAL observations 
             (Table~\ref{tab:observations}).}
  \label{fig:asm}
\end{figure}

As part of a Target of Opportunity programme on
known black hole candidates, 3 INTEGRAL observations of the
region of sky containing \src\ were performed, 
each separated by about one
week (Table~\ref{tab:observations}). The
2--12~keV RXTE All-Sky Monitor (ASM) light curve and hardness ratio
plots of part of the 2003 outburst are shown in Fig.~\ref{fig:asm}
with the times of the INTEGRAL observations indicated. These
took place during the initial $\sim$30~day 
rising part of the outburst, with the third observation close to the
2--12~keV outburst maximum.
The hardness ratio plot indicates that \src\ became softer 
as the intensity increased towards the maximum.

\begin{table}
\caption[]{Observation log. During each observation INTEGRAL 
performed its standard 5 x 5 dither pattern centered close to 
\src\ with a step size of $2\degmark$ and an exposure of 2200~s
at each of $N$ pointings.}
\begin{flushleft}
\begin{tabular}{lcccr}
\hline
\hline\noalign{\smallskip}
Obs & Start Time (UTC) & End Time (UTC) & $N$ & Exp (ks)\\
\noalign{\smallskip\hrule\smallskip}
1 & 2003~Apr~06~15:41 & Apr~07~14:48 & 34 & 74.8 \\
2 & 2003~Apr~14~00:21 & Apr~15~06:13 & 46 & 101.2 \\
3 & 2003~Apr~21~06:24 & Apr~22~14:16 & 46 & 101.2 \\
\noalign{\smallskip\hrule\smallskip}
\end{tabular}
\end{flushleft}
\label{tab:observations}
\end{table}

The data from the 3 observations were processed using the Off-line
Scientific Analysis (OSA) software provided by
the INTEGRAL Science Data Centre (ISDC; Courvoisier et al.~\cite{c:03}). 
This includes  pipelines for the
reduction of INTEGRAL data from all four instruments. The 3 high-energy
instruments use coded masks to provide imaging information. This means that
photons from a source within the field of view (FOV) are
distributed over the detector area in a pattern determined by the
position of the source in the FOV. Source positions and intensities are
determined by matching the observed distribution of counts with those 
produced by the mask modulation. 

\section{Results}

\begin{figure*}
  \hbox{\hspace{0.1cm}
  \includegraphics[height=8.0cm,angle=0]{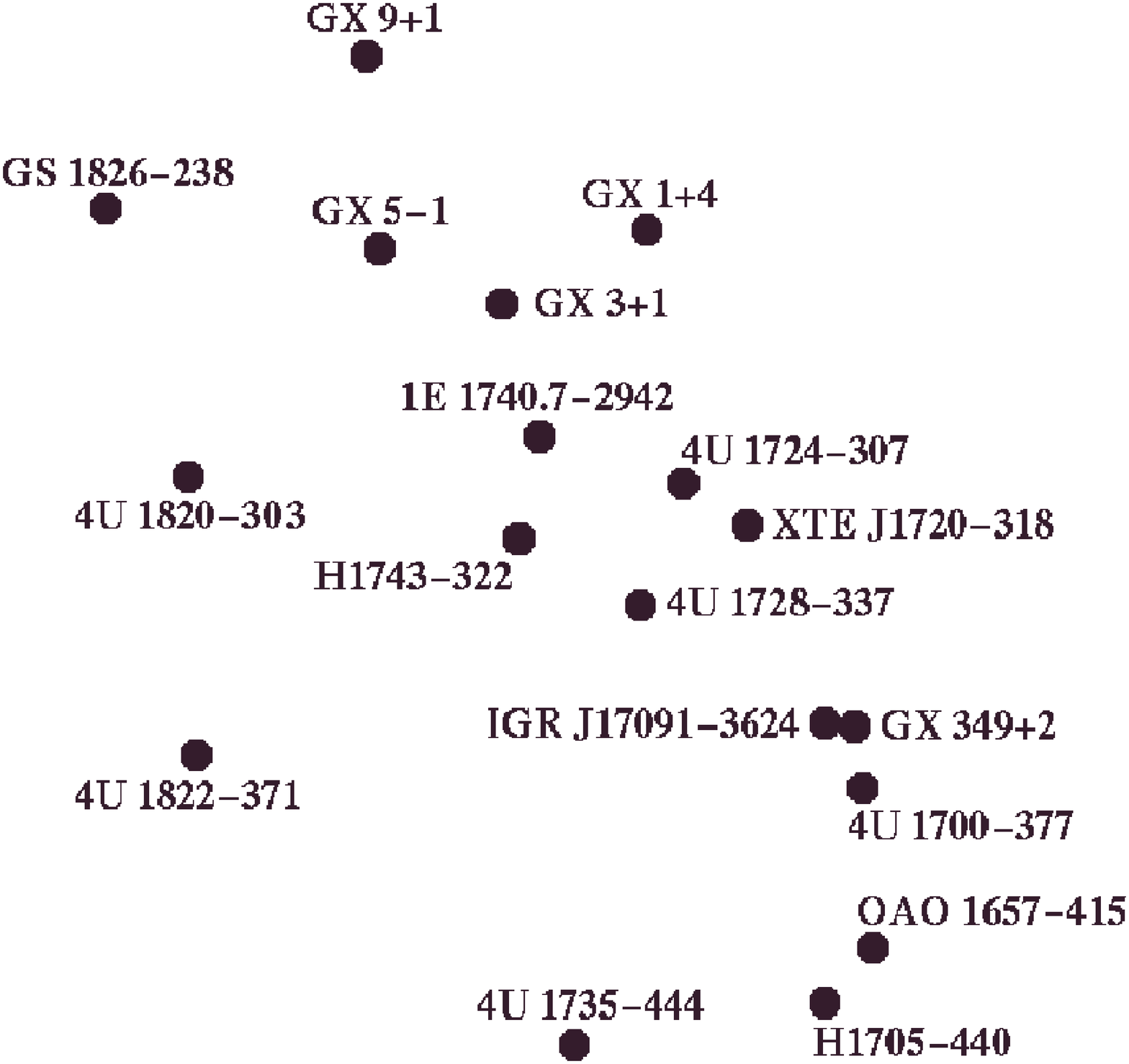}
  \hspace{0.1cm}
  \includegraphics[height=8.0cm,angle=0]{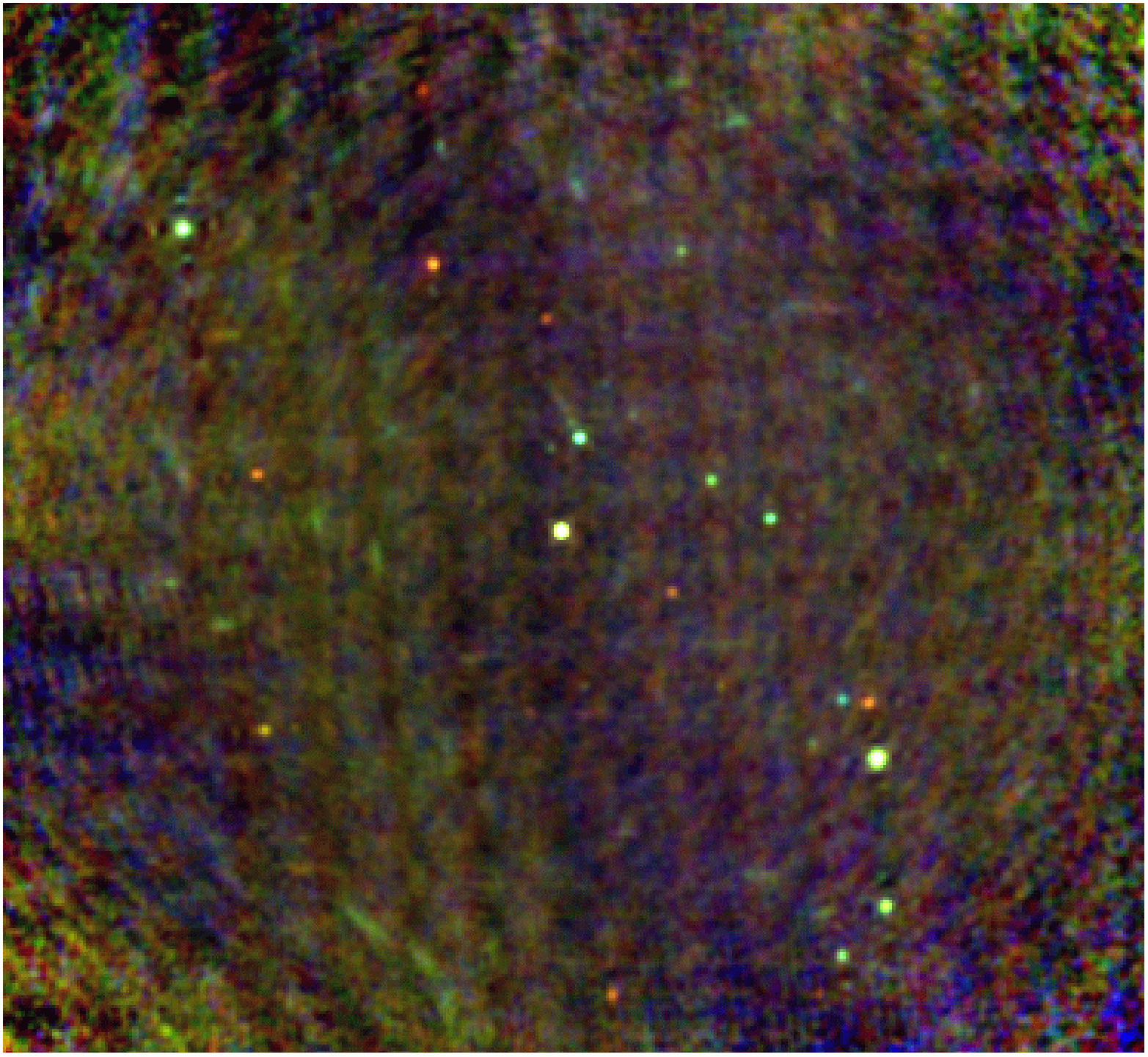}}
  \caption[]{A color coded ISGRI image of a 
             region of sky $\sim$27 x 27$\degmark$ centered close to \src.
             Identifications of the brighter sources are shown in the left 
             panel. The exposure is 277~ks and the 15--40~keV, 40--100~keV,
             and 100--200~keV intensities are displayed in red, green, 
             and blue, respectively. Low-mass X-ray binaries containing 
             neutron stars (e.g., GX\,5$-$1) have ``soft'' spectra and 
             appear as reddish objects, while black hole candidate sources 
             such as 1E\,1740.7$-$2942 appear blue due to 
             their ``harder'' X-ray emission. The 15--200~keV count rate
             from \src\ is $\sim$20 times that of GX\,3+1, one of the
             faintest identified sources in the image.
             The structured background is due to remaining 
             uncertainties in the instrument response modeling.}
  \label{fig:color_image}
\end{figure*}

\begin{figure}
  \hbox{\hspace{0.5cm}
  \includegraphics[width=7.5cm,angle=0]{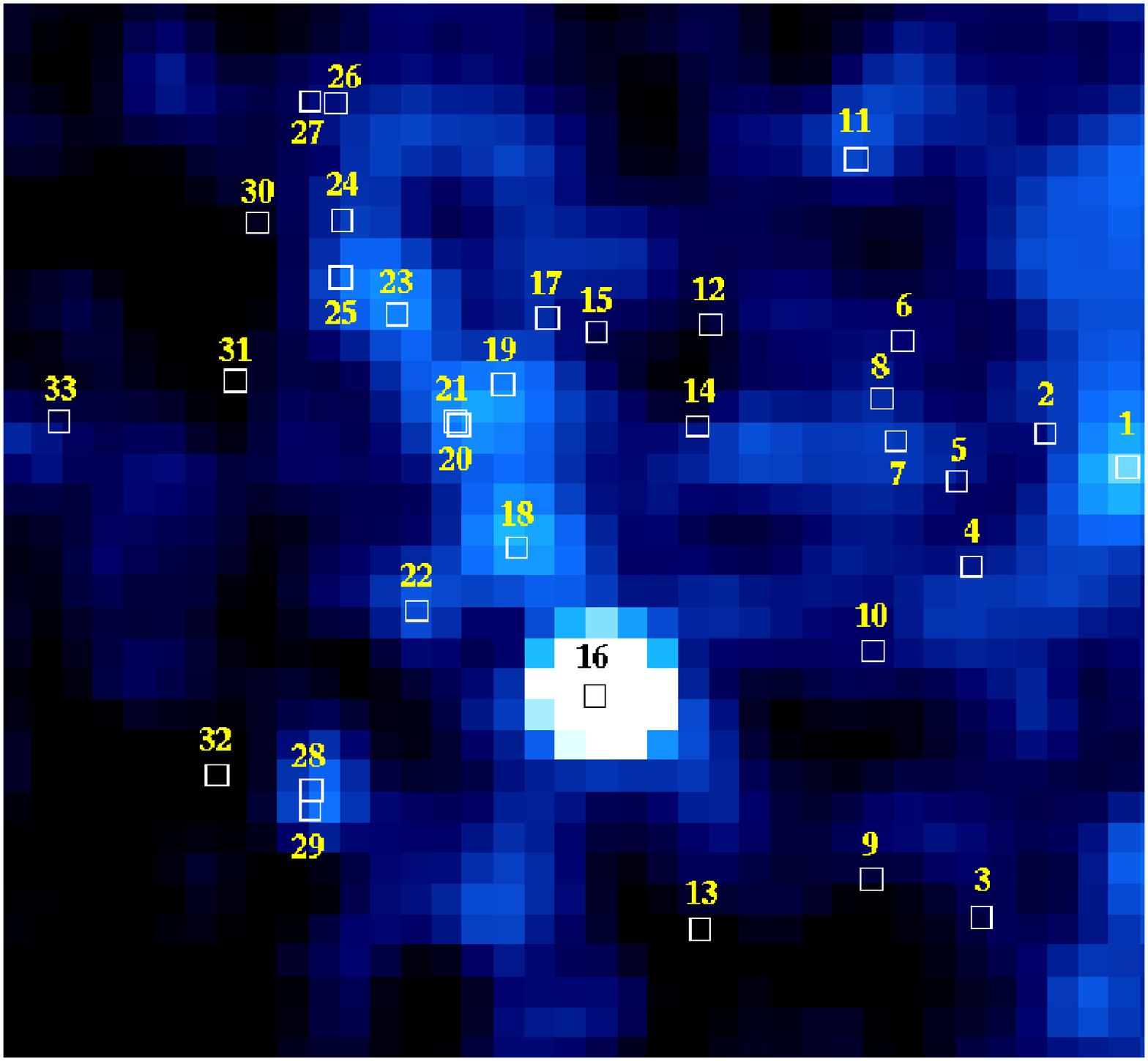}}
  \caption[] {A 15--40~keV ISGRI 3.2$\degmark$ x 3.5$\degmark$ image of
             the diffuse ``tail'' seen in Fig.~\ref{fig:color_image}
             obtained using data from all 3 observations.
             Sources listed in the 0.7--10~keV ASCA
             galactic center catalog (Sakano et al.~\cite{s:02})
             are numbered and identified below.  
             The scaling is linear with the counts in brightest pixel of
             source 16 (1E\,1740.7$-$2942)
             $\sim$10 times more than in that of source 28/29. 
             It is likely that the ``tail'' results from the
             superposition of emission from a number of
             point sources.
 1 = GRS\,1734-292,
 2 = RX\,J1738.4-2901,
 3 = XTE\,J1739-302,
 4 = AX\,J1739.3-2923,
 5 = AX\,J1739.5-2910,
 6 = AX\,J1740.1-2847,
 7 = AX\,J1740.2-2903,
 8 = AX\,J1740.4-2856,
 9 = SAX\,J1740.5-3013,
10 = AX\,J1740.5-2937, 
11 = SLX\,1737-282,
12 = AX\,J1742.5-2845,
13 = AX\,J1742.6-3022,
14 = AX\,J1742.6-2903,
15 = AX\,J1743.9-2846,
16 = 1E\,1740.7-2942,
17 = GRO\,J1744-28,
18 = KS\,1741-293,
19 = GRS\,1741.9-2853,
20 = AX\,J1745.6-2901,
21 = Sgr\,A*,
22 = A\,1742-294,
23 = 1E\,1743.1-2843,
24 = AX\,J1747.0-2828,
25 = AX\,J1747.0-2837,
26 = AX\,J1747.1-2809,
27 = SNR\,G0.9+0.1,
28 = SLX\,1744-299,
29 = SLX\,1744-300,
30 = XTE\,J1748-288,
31 = AX\,J1748.3-2854,
32 = HD\,316341,
33 = SAX\,J1750.8-2900.

} 
  \label{fig:galcen}
\end{figure}

Since \src\ is located close to the galactic centre 
($l = 357\degmark,\,b = -2\degmark)$ where
many bright X-ray sources are located, 
images of the region of sky observed by IBIS 
were produced in order to ensure 
that \src\ could be resolved from any nearby sources and
to study the surrounding field.
ISGRI images were produced in the 15--40~keV,
40--100~keV and 100--200~keV energy ranges by summing 
data from the 126 individual pointings of all 
3 observations. \src\ was clearly detected in all 3 images which
were then combined into a single multi-color image 
(Fig.~\ref{fig:color_image}).
At least 18 sources are clearly evident and can be identified
with well known persistent X-ray binaries or transients such
as IGR\,J17091$-$3624 (Kuulkers et al.~\cite{k:03})
which were active at the time. 
The spectral capability of ISGRI is well demonstrated by
the various colors of different types of sources in 
Fig.~\ref{fig:color_image}. 
Low-mass X-ray binaries with neutron star compact objects such as GX\,5$-$1, 
GX\,9+1, GX\,3+1, GX\,349+2, 4U\,1820$-$303, 4U\,1735$-$444 and
4U\,1728$-$337 appear as reddish objects, consistent with 
their ``soft'' spectra, 
while sources such as 1E\,1740.7$-$2942 and IGR\,\,J17091$-$3624,
which may contain black holes, appear blue due to 
their ``harder'' spectra.   
Closer examination of the ``tail'' extending from 
1E\,1740.7$-$2942 indicates that this is probably unresolved
emission from a number of known X-ray point sources (Fig.~\ref{fig:galcen}).

\begin{figure}
  \hbox{\hspace{0.4cm}
  \includegraphics[height=7.5cm,angle=-90]{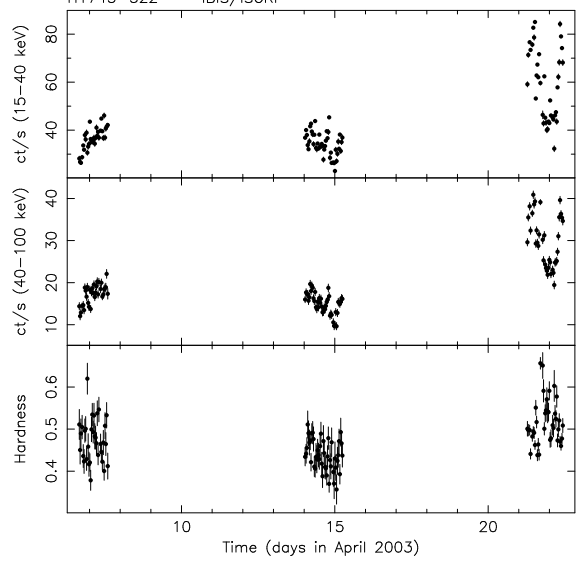}}
  \caption[]{ISGRI 15--40~keV and 40--100~keV background subtracted \src\
             light curves and (40--100~keV/15--40~keV) hardness ratio 
with a binning of 2200~s.}
  \label{fig:lc}
\end{figure}

Figure~\ref{fig:lc} shows ISGRI 15--40~keV and 40--100~keV  
background subtracted
light curves of \src\ with a binning of 2200~s.
The lower panel
shows the hardness ratio (counts in the energy range 
40--100~keV divided by those between 15--40~keV). The intensities
during the first 2 observation are similar, while during the
third observation the source is clearly brighter and more variable.
In contrast to the ASM 2--12~keV hardness ratio (Fig.~\ref{fig:asm}),
the 15--100~keV ISGRI hardness ratio increased slightly as the 
source intensity increased.

Attempts to extract ISGRI spectra of \src\ using OSA v~1.1
were unsuccessful. Following discussions with the ISDC this is likely
due to
a known limitation of the software, especially in crowded fields, 
which will be fixed in a future release. 
We were unable to combine PICsIT data from individual 2200~s
pointings, so these data were not considered further.
Figure~\ref{fig:color_image} was examined
to see if there are any strong sources 
close to \src\ which could cause source confusion in SPI
with its $2\fdg7$ FWHM spatial resolution.
The closest such source is 1E\,1740.7$-$2942 which is $2\fdg5$ away, and
well resolved in SPI images.
No such problems are expected
with JEM-X with its $3\arcmin$~FWHM spatial resolution and so 
the subsequent spectral analysis concentrates on JEM-X and SPI.

\begin{figure}
  \includegraphics[height=8.7cm,angle=-90]{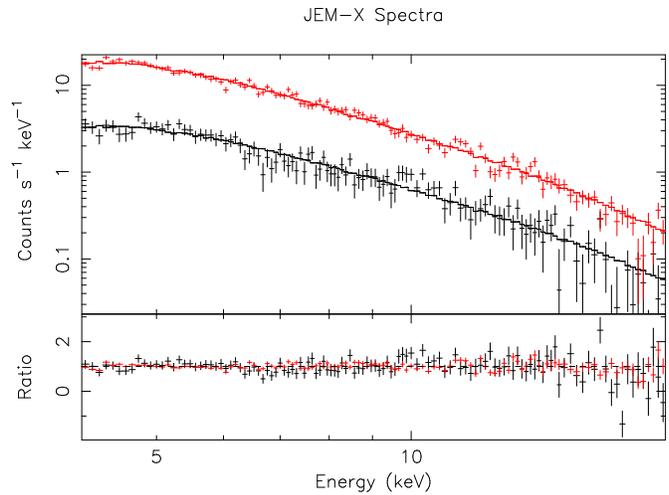}
  \caption[]{JEM-X spectra of \src\ during low and high count rate
             intervals (see text) fitted with absorbed power-law models.}
  \label{fig:jem-x_spectra}
\end{figure}

During parts of the 5 x 5 dither patterns \src\ was outside the 
JEM-X FOV and only JEM-X spectra for the pointings when \src\ was 
close to the center of the FOV were extracted. 
In order to search for any intensity dependent spectral changes in
the JEM-X energy range,
on-axis spectra during low (2003~April~15~00:26~UTC to 01:03)
and high (2003~April~21~14:55~UTC to 15:32) count rate intervals were 
selected (Fig.~\ref{fig:jem-x_spectra}). 
An absorbed power-law model was fit to both spectra to give
$\alpha$ = $3.00\, ^{+0.21} _{-0.07}$ and $3.27 \, ^{+0.08} _{-0.04}$
and $N_H$ $<$$4 \times 10^{22}$~atom~cm$^{-2}$ and
$<$$2 \times 10^{22}$~atom~cm$^{-2}$ for the low and high count rate
intervals, respectively (all spectral uncertainties 
and upper limits are given at 90\%
confidence). Thus, there is some evidence for the source
spectrum softening with increased 4--20~keV intensity,
consistent with the change seen in the 2--12~keV ASM data
(see Fig.~\ref{fig:asm}).
The upper-limits to $N_H$ are consistent
with the absorption of $7 \times 10^{21}$~atom~cm$^{-2}$ in the 
direction of \src\ (Dickey \& Lockman~\cite{d:90}).
In the 5--20~keV energy range the best-fit absorbed fluxes correspond
to 90 and 400~mCrab during the 2 intervals.

A SPI spectrum was created from data obtained
between 2003 April 14 01:11 and April 15 06:17~UTC.
Initial attempts to extract SPI spectra resulted in 
very structured residuals in
which background features were clearly visible. These were also evident
in the image reconstruction process where 3 individual pointings
contributed excessively to the total $\chi^2$
and were excluded from further analysis. The
resulting exposure time is 98.7 ks.
The 25--200~keV SPI spectrum was fitted together with the 
5--20~keV JEM-X spectrum 
obtained during the low-count
rate interval. The JEM-X to
SPI relative nomalisation was left free in the fitting to account for
systematic uncertainties in the instrumental responses and the
non-simultaneity of the spectra.
We note that there were no large changes in hardness ratio
during this observation (Fig.~\ref{fig:lc}). 
The absorbed power-law model
gives $\alpha$ = $2.70 \pm 0.09$ and 
$N_H$ = $(2.5 \, ^{+4.3} _{-2.5}) \times 10^{22}$~atom~cm$^{-2}$ 
with a $\chi ^2$ of 140 for 122 degrees of freedom.
The best-fit
value for the relative JEM-X normalisation is 0.55 and
the 5--200~keV flux corresponds to 210~mCrab.

\begin{figure}
  \includegraphics[height=8.7cm,angle=-90]{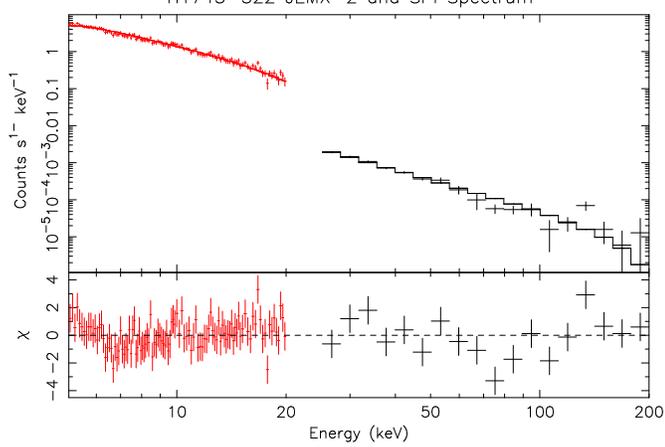}
  \caption[]{The combined 5--200~keV JEM-X (red) and SPI spectrum of \src\ 
             fitted with an absorbed ($\alpha = 2.70$) 
             power-law model. The contributions
             to $\chi$ are shown in the lower panel.}
  \label{fig:jemx_spi_spectrum}
\end{figure}

The SPI spectrum is slightly harder than that from
JEM-X. It is unclear 
whether this is due to uncertainties in data extraction and/or 
instrument calibration, or additional complexity in the source spectrum,
perhaps caused by the presence of a reflection component.
We note that signatures of Compton reflection are apparant in the RXTE PCA
spectrum of Markwardt \& Swank~(\cite{ms:03a}).
As the somewhat structured residuals in 
Fig.~\ref{fig:jemx_spi_spectrum} show, there may be additional
features in the SPI background which are not properly accounted for in the
extraction process.

\section{Discussion}
\label{sect:discussion}

We report on 3 INTEGRAL observations in 2003 April of the poorly
studied black hole candidate \src\ during the rise and close to the
maximum of an outburst. 
Unfortunately, it is not yet possible to extract
an ISGRI spectrum of \src. The combined 5--200~keV 
JEM-X and SPI spectrum was successfully 
fit with an absorbed 
($N_H$ = $(2.5 \, ^{+4.3} _{-2.5}) \times 10^{22}$~atom~cm$^{-2}$)
soft ($\alpha = 2.70 \pm 0.09$) power-law model. 
Based on this spectral shape, and the small
differences in hardness ratio between the 3 observations, the source was in
the canonical high/soft state observed from many 
black hole candidates (see e.g., McClintock \&
Remillard~\cite{mr:03}) during all 3 observations.

Close to the previous outburst maximum, in mid-September 1977, \src\ 
exhibited a 25--180~keV power-law spectrum with $\alpha = 2.6 \pm 0.2$
(Cooke et al.~\cite{c:84}), similar to that measured here.
Six months later the 1--10~keV intensity had fallen by a factor of
10 (Wood et al.~\cite{w:77}), but the intensity at $\sim$100~keV 
had hardly changed (see Fig.~2 of Cooke et al.~\cite{c:84}), consistent
with a hardening of the spectrum with decreasing intensity, as seen
during the 2003 outburst (e.g., Fig.~\ref{fig:asm}) and
the power-law $\alpha$ of $1.49 \pm 0.01$ 
measured by the PCA earlier in the outburst (Markwardt \& Swank~\cite{ms:03a}).
The PCA $N_H$ of $2.4 \times 10^{22}$~atom~cm$^{-2}$
is consistent with that reported here suggesting that the 
absorption did not change strongly, at least during the
early part of the outburst. 
There is no evidence in the INTEGRAL data presented here, nor in the PCA data
of Markwardt \& Swank~(\cite{ms:03a}), for an additional low-energy 
component. However, such a component was detected later 
in the outburst on 2003 May 28 by the PCA (Homan et al.~\cite{h:03}) and 
$may$ have been seen
during the 1977--1978 outburst by HEAO~1 (R. Doxsey 1980, private 
communication in Cooke et al.~\cite{c:84}), but no published spectrum
can be found. A more detailed analysis of the INTEGRAL
data will allow a sensitive search for this component, 
the signature of reflection, and for
the 511~keV annihilation feature observed from the black
hole candidates  1E\,1740.7$-$2942 (Bouchet et al.~\cite{b:91})
and GRS\,1124$-$684 (Goldwurm et al.~\cite{g:92}).

Finally, \src\ is located in a crowded part of the
sky and at least
18 sources are clearly detected in the IBIS FOV. These
are well known accreting X-ray binaries as well as 3 transients 
in outburst. 
These observations very well illustrate the important role that INTEGRAL can
play in the study of bright galactic X-ray sources, especially in
crowded regions of sky where non-imaging instruments such as the
BeppoSAX Phoswich Detection System (15--300~keV; 
Frontera et al. \cite{f:97}) may have difficulties with source 
confusion.

\begin{acknowledgements}

We thank the RXTE instrument teams at MIT and NASA/GSFC for 
providing the ASM light curves. 

\end{acknowledgements}

\end{document}